# A Survey on Software Testing Techniques using Genetic Algorithm

Chayanika Sharma[1], Sangeeta Sabharwal[2], Ritu Sibal[3]

Department of computer Science and Information Technology, University of Delhi, Netaji Subhas Institute of Technology

Azad Hind Fauz Marg, Dwarka, Sector -3, New Delhi - 110078, India

**Abstract**

The overall aim of the software industry is to ensure delivery of high quality software to the end user. To ensure high quality software, it is required to test software. Testing ensures that software meets user specifications and requirements. However, the field of software testing has a number of underlying issues like effective generation of test cases, prioritisation of test cases etc which need to be tackled. These issues demand on effort, time and cost of the testing. Different techniques and methodologies have been proposed for taking care of these issues. Use of evolutionary algorithms for automatic test generation has been an area of interest for many researchers. Genetic Algorithm (GA) is one such form of evolutionary algorithms. In this research paper, we present a survey of GA approach for addressing the various issues encountered during software testing.

## 1. Introduction

Testing is primarily done on software as well as in web for testing client and server architecture. Software testing is one of the major and primary techniques for achieving high quality software. Software testing is done to detect presence of faults, which cause software failure. However, software testing is a time consuming and expensive task [29], [20], [28]. It consumes almost 50% of the software system development resources [3], [20]. Software testing can also be defined as process of verifying and validating software to ensure that software meets the technical as well as business requirements as expected [16].

Verification is done to ensure that the software meets specification and is close to structural testing whereas validation is close to the functional testing and is done by executing software under test (SUT) [18]. Broadly, testing techniques include functional (black box) and structural (white box) testing. Functional testing is based on functional requirements whereas structural testing is done on code itself [13] [10] [24]. Gray box testing is hybrid of white box testing and black box testing [8].

Testing can be done either manually or automatically by using testing tools. It is found that automated software testing is better than manual testing. However, very few test data generation tools are commercially available today [14]. Various techniques have been proposed for generating test data or test cases automatically. Recently, lot of work is being done for test cases generation using soft computing techniques like fuzzy logic, neural networks, GA, genetic programming and evolutionary computation providing keys to the problem areas of software testing.

Evolutionary testing is an emerging methodology for automatically producing high quality test data [10]. GA is well known form of the evolutionary algorithms conceived by John Holland in United States during late sixties [6] [25]. In [21], evolutionary black box testing is also applied on embedded systems to test its functional and non-functional properties. GA has been applied in many optimization problems for generating test plans for functionality testing, feasible test cases and in many other areas [5] [15]. GA has also been used in model based test case generation [3] [23] [26] [27]. In object oriented unit testing as well as in the black box testing, GA is used for automatic generation of test cases [23], [10], [15]. Concerning testing of web applications, many tools, new techniques and methods have been developed to address issues like maintainability, testability, security, performance, correctness and reliability of web application [8]. Web applications are composed of web pages and components and interaction between them executes web servers, HTTP, browser (the client side) and networks. A web page is information viewed on the client side in a single browser window [16]. In [30], user session data of web application is used to generate test cases by applying GA.

In this research paper, a survey of different software testing techniques where GA is efficiently used is presented. This paper is divided into 4 sections. Section 2 describes briefly the working of a GA. In section 3, applications of GA in different types of software testing is described. Section 4 concludes the paper and gives an overview of our future work.

## 2. Genetic Algorithm: A Brief Introduction

In the past, evolutionary algorithms have been applied in many real life problems. GA is one such evolutionary algorithm. GA has emerged as a practical, robust optimization technique and search method. A GA is a search algorithm that is inspired by the way nature evolves species using natural selection of the fittest individuals.

The possible solutions to problem being solved are represented by a population of chromosomes. A chromosome is a string of binary digits and each digit that makes up a chromosome is called a gene. This initial population can be totally random or can be created manually using processes such as greedy algorithm. The pseudo code of a basic algorithm for GA is as follows [6]:-

```
Initialize (population)
Evaluate (population)
While (stopping condition not satisfied)
{
Selection (population)
Crossover (population)
Mutate (population)
Evaluate (population)
}
```

GA uses three operators on its population which are described below:-

- **Selection**: A selection scheme is applied to determine how individuals are chosen for mating based on their fitness. Fitness can be defined as a capability of an individual to survive and reproduce in an environment. Selection generates the new population from the old one, thus starting a new generation. Each chromosome is evaluated in present generation to determine its fitness value. This fitness value is used to select the better chromosomes from the population for the next generation.

- **Crossover or Recombination**: After selection, the crossover operation is applied to the selected chromosomes. It involves swapping of genes or sequence of bits in the string between two individuals. This process is repeated with different parent individuals until the next generation has enough individuals. After crossover, the mutation operator is applied to a randomly selected subset of the population.

- **Mutation**: Mutation alters chromosomes in small ways to introduce new good traits. It is applied to bring diversity in the population.

## 3. Using Genetic Algorithm in Software Testing

In this section we will discuss in detail the applications of GA in different areas of testing like test planning [5], minimization of test cases in regression testing [11], model based testing [3] [23] [26] [27] and web testing [30].

### 3.1 Applications of GA in White Box Testing

Structural testing can be done in the form of data flow testing or path testing. Path testing involves generating a set of paths that will cover every branch in the program and finding the set of test cases that will execute every path in this set of program path [16] [18]. In data flow testing, the focus is on the points at which variables receive values and the points at which these values are used [2]. Next, we will discuss briefly some of the research work regarding the applications of GA in white box testing.

3.1.1 Data Flow Testing

*M.R. Girgis*
Girgis [7] has proposed a structural oriented automatic test data generation technique that uses a GA guided by the data flow dependencies in the program to fulfil the all-uses criterion. The program to be tested is converted into a Control Flow Graph (CFG) where each node represents a block in a program and the edges of the flow graph depict the control flow of the statements.

Variables in a program under test are divided into 'c-uses' and 'p-uses' variables. c-uses variables are those which are used in computations or as a predicates in a program whereas p-uses variables are associated with edges of the flow graph. In order to fulfil the all-uses criteria, the def-clear path (a path containing no new definition of a current variable) from each definition of a variable to each use of that variable need to be determined. To find out the set of paths satisfying all-uses criteria, it is necessary to determine def c-use(dcu) and def p-use(dpu) of a variable i.e. the def- clear paths to their c-use at node i and def-clear paths to their p-use at edge (i, j).

Using the location of a variable defs and uses in a program under test, combined with the 'Basic state reach algorithm', the sets dcu(i) and dpu(i,j) are determined. From the 'Basic state reach algorithm' two sets reach (i) and avail (i) are determined where reach (i) is the set of all variable defs that "reach" node i and set avail (i) is the set of all "available" variable defs at node i i.e. the union of set of global defs at node i and the set of all defs that reach this node.

$$dcu(i): \ reach(i) \cap c-use(i) \quad (1)$$

$$dpu(i, j): \ avail(i) \cap p-use(i, j) \quad (2)$$

List of all dcu and dpu sets in the procedure calls satisfying the all-uses criterion are determined along with killing nodes (nodes containing other definition of a variable in a current path) that must not be included in the current path. In this approach, GA accepts instrumented version of the program under test, the list of def-use sets to be covered, the number of input variables, and the domain and the precision of each input variables as an input. A binary vector is used to represent a chromosome. The length of the input is determined by the domain and the precision. The domain is represented by $D_i = [a_i, b_i]$ where each variable in a program takes values from the range $[a_i, b_i]$. Each domain $D_i$ should be cut into $(b_i - a_i).10^{d_i}$ equal size ranges if decimal places $d_i$ is desired for a variable to achieve precision. If $m_i$ is an integer denoting the length of a chromosome or a string such that $(b_i - a_i).10^{d_i} \leq 2^{m_i} - 1$ then a binary string denoting a variable of length $m_i$ fulfil the precision requirement. The mapping from the binary string i, into a real number from the range $[a_i, b_i]$ is performed by the following formula:-

$$x_i \quad a_i + x_i' \cdot \frac{b_i - a_i}{2^{m_i} - 1} \quad (3)$$

Where, $x_i'$ represents the decimal value of the binary string i.

$$x_i \quad a_i + \text{int}(x_i' \cdot \frac{b_i - a_i}{2^{m_i} - 1}) \quad (4)$$

By applying $d_i = 0$, the above formula can be used to map binary string i into an integer number from the range $[a_i, b_i]$. Each chromosome represents a test case for a program which is represented by a binary string of specified length. Each chromosome is then represented by a decimal number by using (3) or (4).

The fitness value *eval* $(v_i)$ for each chromosome $v_i$ ($i = 1,...., pop\_size$) is calculated as follows:-

$$eval\ (v_i) = \frac{no.of\ def - use\ paths\ covered\ by\ v_i}{total\ no.of\ def - use\ paths} \quad (5)$$

A test case, $v_i$ is effective if its fitness value *eval* $(v_i) > 0$. Each test case or chromosome is evaluated and the program is executed to record the def – use paths in the program that are covered by the test cases as its input.

All the test cases are selected with effective eval $(v_i)$ or good fitness value. Selection is done by roulette wheel selection and proposed random selection method. The effective test cases then become parents of the new population. If none is effective then all the individuals are chosen as the parents.

*Discussion*
The test case generation by the proposed GA is more effective as compared to the random testing technique. The proposed selection method generates better results than the roulette wheel selection method. However, the proposed selection method has not been evaluated and compared with other selection methods like stochastic, uniform and tournament selection methods.

3.1.2 Path Testing

*P.R Srivastava et al.*
In [20], P.R. Srivastava and Tai have presented a method for optimizing software testing efficiency by identifying the most critical path clusters in a program. The SUT is converted into a CFG. Weights are assigned to the edges of the CFG by applying 80-20 rule. 80 percentage of weight of incoming credit is given to loops and branches and the remaining 20 percentage of incoming credit is given to the edges in sequential path. The summation of weights along the edges comprising a path determines criticality of path. Higher the summation more critical is path and therefore must be tested before other paths. In this way by identifying most critical paths that must be tested first, testing efficiency is increased.

Another test generation approach proposed by P.R Srivastava is based on path coverage testing [19]. The test data is generated for Resource Request algorithm using Ant Colony Optimization algorithm (ACO) and GA. *Resource request algorithm* is deadlock avoidance algorithm used for resource allocation by operating system to the processes in execution cycle [10]. The ACO algorithm is inspired from behaviour of real ants where ants find closest possible route to a food source or destination. The ants generate chemical substance called pheromones which helps ants to follow the path. The pheromone content increases as more ants follow the trail. The possible paths of CFG are generated having maximum number of nodes. Using ACO, optimized path ensuring safety sequence in resource request algorithm is generated covering all edges of CFG.

Using GA, suitable test data set is generated which covers the need for each process. The backbone of genetic process is the fitness function which counts number of times a particular data enters and continues the resource request algorithm. Higher the value of count, higher is chances of avoiding a deadlock. The test data with higher values of count is taken and genetic crossover and mutation is applied to yield better results. Simultaneously, poor test data is removed each time.

*Discussion*
The experimental results shows that success rate of ACO are much better than GA. In weighted CFG approach [20], experiments were done on small examples and need to be done on larger commercial examples. Moreover, method can be further improved.

*Dr. Velur et al.*

The approach for test cases generation from directed graph has been proposed by Dr. Velur [29]. In this work, directed graph of intermediate states of system under test is created to exhibit the expected behaviour of system as shown in Fig 1.

A directed graph is represented by G = [V, E], where V represents state or vertices and edges represents flow of control. Thereafter, a graph containing n nodes is represented by an incidence matrix of order n * n where an entry '1' in the matrix represents edge between nodes and '0' represents no edge or connection between them (see Fig 1). By using the nodes of graph as the base population, pair of nodes are generated which are selected as parents by applying the dual graph generation technique proposed as shown in Fig 2.

The population is initialized by random selection of graphs of size 43 and 250 individuals are generated. The tournament selection method is used, where two individuals are chosen randomly and individual with the maximum fitness is chosen for crossover. Fitness is calculated by using the '*Current maximum clique algorithm*' and *Approximation algorithm*'. Fitness is assigned by finding the clique of size 5 and the number of independent sets of size 5 in the population which comprise of number of graphs in the population. The graph with 0 fitness value indicates the clique of size 0 and no independent sets of size 5 in the graph. Nodes which are already visited are discarded and GA cycle continues till all the nodes of the graph are visited once.

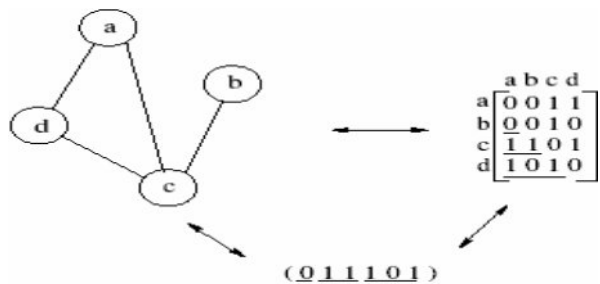

Fig. 1 Graph representation as a binary string [29]

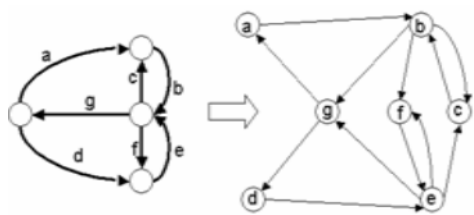

Fig. 2 Dual graph generation [29]

The graph is first converted into a binary string as shown in Fig 1. Next, the arcs of an original graph are converted into nodes as shown in Fig 2. For example, if an edge1 is an incoming to some node and the edge 2 is outgoing edge for the same node then an edge is created from edge 1 to edge 2 which acts as nodes in corresponding dual graph. The dual graph is then eulerized by duplicating the arcs for balancing the node polarities. As dual graph is traversed, all possible two links combination in dual graph for example bc, bf, bg ... are noted down. All the dual combinations are then encoded in 0 and 1 format as genetic population.

*Discussion*

This technique will be more suitable for network testing and system testing where predictive model based tests are not optimized to generate the outputs. The approach uses tournament selection method only. This approach has not been compared with new proposed approaches for generating test cases from the CFG. Moreover, the effectiveness of the fitness evaluation criteria has not been justified.

*Maha Alzabidi et al*.

Maha Alzabidi [14] has proposed automatic structural test case design using evolutionary testing. Software structural testing is done by taking path coverage for testing. For path testing, CFG is used in their work for representing a program where the nodes are the basic blocks and the edges between the nodes indicate the flow of the program. Meaningful paths are extracted from CFG and are selected as a target path for testing. Test cases are generated to trace the new path which leads to the target path. The test result is evaluated to determine that the testing objective criteria are satisfied by executing the selected path.

The fitness function is named as a Shifted–Modified-Similarity (SMS) which is a modification to the hamming distance. The symmetric difference or hamming distance is calculated for cascading edges for target path and current path. Similarities are then normalized and summed associated with a weighting factor. This value is used as an objective function to evaluate the individuals in the population.

*Discussion*

The approach improves the fitness function. Performance of different GA parameters is studied in this paper. Parameters have been applied on different test programs and results have shown that double crossover is more effective than single crossover applied on a test program. The approach is applied on the small programs but has not been evaluated on the complex programs involving loops, arrays and linked lists using different data types.

*Jose Carlos et al.*

The structural testing of object oriented programs requires traversing the complex control flow paths, resulting in the complex test cases generation which defines the elaborate state scenarios. Jose has proposed the methodology for evaluating the quality of both feasible and unfeasible test cases for structural oriented unit testing of object oriented java programs [10]. The test cases that are terminated with a call to a method and are completely executed are termed as feasible test cases whereas the test cases which abort prematurely are termed as unfeasible test cases. In this work, the test cases are represented as strongly typed genetic programming (STGP) individuals where each individual contains number of STGP trees equal to the number of arguments in the current method or method under test (MUT). The tree is traversed by depth first traversal algorithm which generates the sequence of method calls or scenarios.

Traversing the trees by depth first traversal generates a linear sequence of computations or method call sequence (MCS) of MUT which is represented by a CFG as shown in Fig 3. The nodes in a CFG are assigned the weights and the fitness of test cases is computed. The fitness of feasible test case is computed on the basis of number of times a particular CFG node was exercised by the test cases of previous generations whereas unfeasible test cases are measured by the method calls that threw the exceptions or distance between the runtime exception indexes which results in prematurely termination of a test case. The CFG nodes weights are evaluated at the beginning of every generation according to formula proposed by Carlos:-

$$W_{ni} = (\alpha W_{ni})(\frac{hitC_{ni}}{|T|}+1)(\frac{\sum_{x\varepsilon}N_s^{ni}W_x}{|N_s^{ni}|*\frac{W_{init}}{2}}) \quad (6)$$

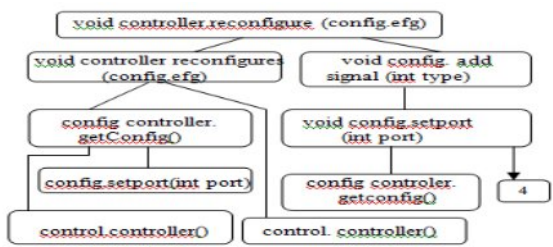

Fig. 3 Example of a STGP tree and corresponding MCS [10]

Where the $hitC_{ni}$ parameter is the "Hit count" representing the number of times a CFG node was traversed by the test cases of the previous versions, T represents the set of test cases in previous generation and α represents weight decrease constant value which ranges from 0 to 1. With this approach unfeasible test cases are considered at certain stages of evolutionary testing, thereby enhancing the diversity and full structural coverage.

*Discussion*

The weights of the CFG are dynamically revaluated each generation. The technique finds a good balance between intensification and diversification of the search by fine tuning the evolutionary operators.

*Bo Zhang and Chen Wang*

Bo zhang and Chen wang [2] used simulated anneal algorithm into GA to generate test data for path testing. A simulated annealing algorithm is inspired by the annealing of metals. In this method, solid is heated from high temperature and then cooled down slowly to maintain thermodynamic equilibrium of system.

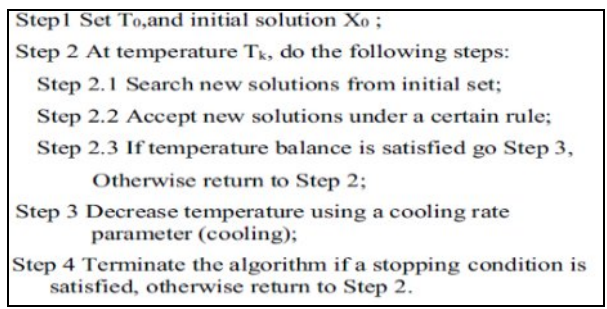

Fig. 4 Simulated Annealing Algorithm [2]

The steps in simulated annealing algorithm are shown in Fig 4. The Adaptive Genetic Simulated Annealing Algorithm (AGSAA) is proposed by Zhang to automatically generate test data. The CFG is used for path coverage testing. The following section shows the fitness function, crossover, mutation and other modifications applied in the GA procedure and elements by Bo Zhang.

Fitness Function: - The fitness function used is named SIMILARITY proposed by Lin and Yeh [9]. The SIMILARITY function is modification to hamming distance. It is used to get distance between two paths. The hamming distance is derived from the symmetric difference in set theory. As stated in [9], symmetric difference between two sets α and β is denoted by $\alpha \oplus \beta$. The symmetric difference between two sets α and β is the set containing the elements either in α or β but not in both. The Bo Zhang expressed the SIMILARITY as stated in equation 7.

$$SIMILARITY_{i,j} = M^1_{i,j} \times W_1 + M^2_{i,j} \times W_2 + \ldots + M^n_{i,j} \times W_n$$

Where,
$M^n_{i,j} = 1 - |S^n_i \oplus S^n_j| / |S^n_i \cup S^n_j|$
$W_n = W_n \times |S^{n-1}_i|$
$S^n_i \oplus S^n_j = (S^n_i \cup S^n_j) - (S^n_i \cap S^n_j)$
$W_1 = 1$

(7)

Using the target path and the current path in the CFG, the fitness function is calculated.

Adaptive Selection: - The adaptive power selection strategy is used by Bo Zhang and Chen Wang. The power function used is shown as:-

$$P_s = \{Y_i = X_j(t)\} = n(X_j(t))J^{a(t)}(X_j(t)) / \sum_{K=1}^{N} J^{a(t)}(X_k(t)) \quad (8)$$

Where, $Y_1, Y_2, \ldots, Y_n$ is new population. $P_s = X_j(t)$ is the probability of selection for $X_j(t)$. $N(X_j(t))$ is number of $X_j(t)$ in current population, $J(X_j(t))$ is fitness of $X_j(t)$, $a(t)$ is a monotonously increasing sequence of positive real and $\sum J(X_k(t))$ is sum of all individuals in the current population.

Elitist Preservation:-The good individuals having good fitness value are protected from being modified.

Adaptive Crossover and Mutation: - In general, GA uses constant crossover and mutation probability but in this work crossover probability and mutation probability changes are according to the fitness value. If fitness value of parent is bigger than average fitness value, probability of crossover is smaller and the parent will be protected from being modified whereas when fitness value is smaller than average fitness value, probability of crossover is large and parent will be died out. For random number r where r Є [0, 1], if r < crossover probability, then parent individuals are selected for crossover. In mutation, the good fitness chromosomes are considered to have smaller mutation probability while bad chromosomes have high mutation probability.

Simulated Annealing: -Simulated annealing is used to decide whether a chromosome is better than the original one and based on those criteria chromosome is accepted or rejected.

*Discussion*
As a case study, Zhang used triangle classification problem for the experiment. The target path is selected from the structural code. In the experiment, initial population size is 100 individuals. Maximum number of generations is 20. The results show that AGSAA performs better than GA in terms of covering the objective path quickly and the rate of coverage.

**3.2 Applications of GA in Black Box Testing**

In functional or black box testing, a program is considered to be a function that maps values from its input domain to values in the output range [18]. In other words, black box testing first concentrates on test to pass and then test to fail. This section describes work on black box testing for test case generation using GA.

3.2.1 Functional Testing

*Francisca Eanuelle et al.*
Francisca Eanuelle [5] has presented GA based technique to generate good test plans for functionality testing in an unbiased manner to avoid the experts interference. The motivation behind this work is to prove that the GA is able to generate good test plan although the best sequence of test plan is unknown.

The test plan or test sequence totally relies on the experts or the people who understand the application well. The emphasis is given on the fact that "*an error in a program is not necessarily due to the last operation executed by the user but may have been due to a sequence of previously executed operations that leads an application in an inconsistent state*". In other words, as a sequence of operations is executed, the state of inconsistency is non-decreasing or a problem in a software application is directly proportional to the level of inconsistency of the state in which application is. In this work, the operation of large granularity has been chosen so that the sequence of operation that leads application to inconsistent state can be identified. The transitions of an operation $l_i$ yield a new operation $l_{i+1}$ which leads the system into a new state as shown in Table 1.

As shown in Table 1, the objective is to find the sequence of operations which leads the system in an inconsistent state. Fitness value for the Table 1 is calculated as:-

$$f_p \quad \sum_{i}^{k-1} t(l_i \rightarrow l_{i+1}) \quad (9)$$

Table 1: Representation of the assigned values for the inconsistency added by each transition for instance $t(l_2 \rightarrow l_3) = v_{2,3}$ [5]

|       | $l_1$     | $l_2$     | $l_3$     | ……    | $l_n$     |
|-------|-----------|-----------|-----------|-------|-----------|
| $l_1$ | $v_{1,1}$ | $v_{1,2}$ | $v_{1,3}$ | …… | $v_{1,n}$ |
| $l_2$ | $v_{2,1}$ | $v_{2,2}$ | $v_{2,3}$ | …… | $v_{2,n}$ |
| $l_3$ | $v_{3,1}$ | $v_{3,2}$ | $v_{3,3}$ | …… | $v_{3,n}$ |
| :     | :         | :         | :         |       |           |
| $l_n$ | $v_{n,1}$ | $v_{n,2}$ | $v_{n,3}$ | …… | $v_{n,n}$ |

Where p = $l_1, l_2,...., l_k$ is a test plan or sequence of operations and t is a transition function for converting one operation $l_i$ to the next operation $l_{i+1}$ in a sequence or in a new state. Larger the value of fitness function, better the sequence is considered which is likely to take the application to an inconsistent state. The GA is applied on the table of size 30*30 with randomly generated transitions values as shown in Table 1. The results have shown that the GA improves the quality of the test plans.

*Discussion*

The fitness function is defined which is used to determine the inconsistency of the application. In this work, a test plan having highest contribution to the inconsistency of the application is considered as a good test plan. This approach can eliminate the bias from the plans generated by experts. A technique based on GA proposed by Francisca, generates good test plans in an unbiased way but this requires computer applications to be tested more thoroughly. The approach does not use the structure of the application or the program flow. Moreover, approach does not deal with data and focus is only at the level of macro operations in functional testing.

*Ruilian Zhao et al.*

In [22], Zhao used the neural network and GA for the functional testing. Neural network is used to create a model that can be taken as a function substitute for the SUT. The emphasis is given on the outputs which exhibit the important features of SUT than inputs. In that case, test cases should be generated from the output domain rather than input domain. The feed forward neural network and back propagation training algorithm is used for creating a model. Neural network is trained by simulating the SUT. The outputs generated from the created model are fed to the GA which is used to find the corresponding inputs so that automation of test cases generation from output domain is completed.

In this paper, inputs to the GA are the function model generated from neural network, number of input variables n, range of input variables i.e. upper [n] and lower [n], population size, maximum iteration number, goal output g, maximum fitness function $f_{max}$, crossover probability and mutation probability. The fitness function is defined as:-

$$f = \begin{cases} \frac{1}{|c-g|} & c \neq g \\ f_{max} & |c-g| \leq 10^{-8} \end{cases} \quad (10)$$

Where c is the actual output and the g is the goal output of the SUT. Population is evaluated by applying GA operations such as reproduction, crossover and mutation. The current individuals generated are considered as test inputs if the fitness value reaches or exceeds $f_{max}$.

$$\begin{aligned} y_1 &= x_1 - r(x_2 - x_1) \\ y_2 &= x_1 + r(x_2 - x_1) \\ y_3 &= x_2 + r(x_2 - x_1) \\ y_4 &= (1-r)x_{min} + rx_1 \\ y_5 &= (1-r)x_{max} + rx_2 \end{aligned} \quad (11)$$

Zhao has proposed new strategy for the crossover operation as shown in equation 11. In equation 12, $x_1$ and $x_2$ are the chosen parent individuals and $y_1, y_2, y_3, y_4, y_5$ are new individuals by applying crossover operation and r is a random number generated in (0, 1).

The difference between goal output and actual output of SUT using neural network is used for calculating fitness value of the individuals in the population. If fitness value exceeds or reaches the maximum fitness value, then search stops and the current individual is taken as the test inputs for the corresponding outputs.

*Discussion*

The test cases are generated from the output domain. Results have shown that this approach can generate test cases from output domain with high efficiency. The experiments were conducted only on small programs. The effectiveness of this approach can be further evaluated on large size programs. The actual outputs of created model are closed to the correct outputs. To minimize this difference, fitness function can also improve.

### 3.2.2 Mutation Testing

*Mark Last et. al.*

Mark Last [13] used the fuzzy based extension of GA (FAexGA) approach for test case generation. The aim is to find minimal set of test cases that are likely to expose faults using mutated versions of the original program. In FAexGA approach, crossover probability varies according to the age intervals assigned during lifetime. The crossover probability of young and old individuals is assigned low while for other age interval this probability is high. The very young offsprings crossover probability is low thus enabling exploration capability. Old offsprings have also less crossover probability and eventually dying out would help avoiding a local optimum or premature convergence. On the other hand, middle-age offsprings are frequently used for crossover operation.

Fuzzy logic controller (FLC) is used for determining probability of crossover. The FLC state variables include the age and lifetime of chromosomes (parents). The emphasis of this work is on the exploration and exploitation of individuals. The fuzzification interface of FLC includes variables that determine the age of an offspring. FLC assigns every parent values Young or Middle-age or Old. These values determine the membership for each rule in FLC rule base. The fuzzification interface of FLC defines for each parents the truth value of being Young, middle-age and old as shown in Table 2.

Table 2: M. Last's fuzzy rule for crossover probability [13]

| Parent 1 / Parent 2 | "Young" | "Middle-age" | "old" |
|---|---|---|---|
| "Young" | Low | Medium | High |
| "Middle-age" | Medium | High | Medium |
| "old" | Low | Medium | Low |

The fuzzy rule base used in this experiment is presented in Table2. Each cell defines a single fuzzy rule. For example, "If Parent 1 is *old* and Parent 2 is *old* then crossover probability is *Low*". The centre of gravity (COG) is used as a defuzzification method which computes crisp value for the crossover probability based on values of the linguistic labels as shown in Table 2.

The test cases relate to the inputs of tested software and are represented as a vector of binary or continuous values. The test cases are initialized randomly in the search space of possible input values. Genetic operators are applied and the test cases are evaluated based on the fault – exposing - capability using mutated versions of original program.

The Boolean expression composed of 100 Boolean attributes and three logical operators: AND, OR and NOT (correct expression) is taken as the case study. The expression was generated randomly and to define an evaluation function for each test case an erroneous expression is generated. The chromosomes are 1-dimensional binary strings of 100 bit length. The value of the evaluation function F is calculated as follows:-

$$F(T) = \begin{cases} 1, & \text{if } Eval\_Correct \neq Eval\_Erroneous(T) \\ 0, & \text{respectively} \end{cases} \quad (12)$$

Where T is a 100 bit 1-dimensional binary chromosome representing a single test case and Eval_Correct (T) or Eval_Erroneous (T) are binary results of applying chromosome T to the correct or erroneous expression.

*Discussion*
FAexGA has not been evaluated on the real programs. Moreover, sophisticated and continuous evaluation functions need to be developed.

3.2.3 Regression Testing

*Liang You and YanSheng Lu*
In [11], redundant test cases in the regression test suite are deleted and the total running time of remaining test cases are minimized by applying GA. The satisfaction matrix $S_{ij}$ is used to represent relationship between requirements and the test cases. The rows in the satisfaction matrix represent requirements and column represents test cases. $S_{ij} = 1$ represents that $j^{th}$ test case $t_j$ satisfy the $i^{th}$ requirement else $S_{ij} = 0$. The time aware regression testing reduction problem is defined as:-

$$Minimize \sum_{j\ 1}^{n} C_j x_j \quad (13)$$

The fitness function shown in equation 14 and represents total running time of remaining test cases after eliminating redundant test cases. $C_j$ represents the running time of test case $T_j$, $x_j$ is the vector of the test case $T_j$. $x_j = 1$ represents that test case $T_j$ exists in $T_{min}$ and $x_j = 0$ represents that test case $T_j$ does not exists in $T_{min}$. If $T_1 = \{t_1, t_3, t_5\}$, then Require $(T_1) = \{r_1, r_2, r_3, r_4\}$ = R = Require (T). Using greedy algorithm, $T_{min} = T_1 = \{t_1, t_3, t_5\}$. $T_{min}$ is the minimal regression testing suite consisting of seven test cases i.e. T [7]. X = $\{x_1, x_2,...., x_n\}$ is the reduction regression testing suite $T_{min}$. For example, X= $\{1, 0, 1, 0, 1, 0, 0\}$ represents $T_{min} = \{t_1, t_3, t_5\}$. The X = $\{x_1, x_2,...., x_n\}$ is used as a bit string to represent chromosome X. The repair operator is used to transform infeasible solution into feasible solution. The repair operator, crossover and mutation are same as CHU's [17] GA. In CHU's GA, uniform crossover operator is used and mutation rate equals to two bits per string.

$$R_1 = (1 - \frac{number\ of\ reduced\ test\ suite}{number\ of\ unreduced\ test\ suite}) * 100\%$$

$$R_2 = (1 - \frac{total\ running\ time\ of\ reduced\ test\ suite}{total\ running\ time\ of\ unreduced\ test\ suite}) * 100\%$$

(14)

In equation 14, R1 represents reduction rate of number of test cases and R2 represents reduction rate of total running time of test cases. The paper compares GA based reduction with vector based reduction on all test cases (VA).

*Conclusion*
Results show that GA is better than VA reduction strategy in 7 out of 8 case studies and the saving time is greater than saving size of reduced test cases i.e. R2 > R1. Thus, GA is an effective technique for minimizing test cases in a test suite.

### 3.2.4 Model Based Testing

*Chartchai Doungsa et al.*

State diagrams or state chart diagrams are used to help the developer better understand any complex functionality or depict the dynamic behaviour of the entire system, or a sub-system, or even a single object in a system. GA can be used to generate the test data using UML state chart diagram as described by Doungsa [3]. Sometimes after coding developers don't have time to test the software. Generating test cases from UML state chart diagram can solve this problem by generating them before the coding. Then the test cases can be generated as per the specifications of the software. Specifications can be in the form of UML diagrams, formal language specifications or natural language description.

Sequence of triggers for UML state diagram can be used as a chromosome. The sequence of triggers is an input for the state diagram which acts as test cases for a program to be tested. Each trigger is examined to check for the transitions which lead to a new state. Each trigger checks for state and transition coverage. If the trigger can generate new state from the current state then next trigger is checked. If the trigger in a sequence cannot generate a new state then tracing for the state coverage will be stopped and the state and the transition coverage are recorded without taking rest of the sequence to consider. As each trigger is traced, new states and transitions are recorded. The process continues for fixed number of generations until all the states and transitions are covered. Fitness of the chromosome is evaluated by using objective function as follows:

$$a\,W + b\,X + c\,Y + Z \qquad (15)$$

Where, a, b, c are constant value and a = 0 when there is no guard condition in selected transition. W are a number of states in test cases where value of attribute in that state make guard condition to be true. X is a number of transitions which is covered by this test but have not been covered by previous test set. Y is a number of states that can be reached by test case to reachable transition source. Z is a number of state and path coverage for the test case. Test cases are selected based on their fitness function. Test case with best fitness value is selected as parents. Based on the fitness function the selection operator is used to apply crossover and mutation operator to the sequence of triggers. Crossover operator is then applied to the sequence of triggers. This operator then generates new states and transitions. After a new generation is created, UML state diagram is then executed again to check for new chromosomes.

*Discussion*

In this work, test cases are generated from UML state diagram so that test data can be generated before coding. The effectiveness of test cases generated from the proposed fitness function is not evaluated with other test case generation techniques from the UML.

*Sangeeta Sabharwal et. al.*

In this work, software testing efficiency is optimized by identifying critical path clusters [26]. The test case scenarios are derived from activity diagram. The activity diagram is converted into CFG where each node represents an activity and the edges of the flow graph depict the control flow of the activities. Path testing involves generating a set of paths that will cover every branch in the program and finding the set of test case scenarios that will traverse every activity in these scenarios. It may be very tedious expensive and time consuming to achieve this goal due to various reasons. For example, there can exist infinite paths when a CFG has loops.

In this approach, critical path are identified that must be tested first using the concept of information flow (IF) metric and GA. The IF metric is adopted in this work for calculating the IF complexity associated with each node of the activity diagram. According to the basic IF model, IF metric are applied to the components of system design. In this work, the component is taken as a node in the CFG. The IF is calculated for each node of CFG. For example, IF of node A i.e. IF (A) is calculated using equation given below:-

$$IF(A) \quad [FANIN(A) \times FANOUT(A)] \qquad (16)$$

Where FANIN (A) is a count of number of other nodes that can call, or pass control to node A and FANOUT (A) is a number of nodes that are called by node A. IF is calculated for each node of a CFG. The weighted nodes in the path are summed together and the complexity of each path is calculated.

In [27], to take care of the software requirements change and to improve software testing efficiency a stack based approach is adopted for assigning weights to the nodes of an activity diagram and state chart diagram. The nodes of CFG and intermediate graph of state chart diagram i.e. state dependency graph (SDG) are prioritized using stack based memory allocation approach and IF metrics.

In the stack based memory allocation approach, data or info is pushed or popped only at one end called top of stack. The stack uses last in first out (LIFO) approach. Node pushed first is removed last from the stack. The top is incremented when node is inserted and decremented when node is deleted. In the proposed technique, each node of CFG or SDG is assigned a weight w based on number of operations to access element in the stack. To access or modify the node (data), all the data above it is popped. Higher the number of operations required to access the node, higher is the weight or complexity of the node. If the weight of the node or number of operations to access the node increases, the cost of modifying the node also increases.

The IF is calculated for each node of a CFG and SDG. The weighted nodes in the path are summed together and the complexity of each path is calculated. Therefore, the sum of the weight of a node by stack based weight assignment approach and IF complexity contributes to the total weight of a node of CFG and SDG. The fitness value of each chromosome is calculated by using the formula given below:-

$$F \quad \sum_{i\ 1}^{n} w_i \tag{17}$$

Where, $w_i$ is weight of $i^{th}$ node in a path under consideration and n is number of nodes in a current path. Weight of $i^{th}$ node is the sum of IF complexity and stack based complexity given by equation given below.

$$w_i \quad IF(i) + STACKBASED\,WEIGHT(i) \tag{18}$$

*Discussion*

An approach is proposed for identifying the test path that must be tested first. Test paths or scenarios are derived from activity diagram. The approach makes use of IF model and GA to find path to be tested first.

*3.2.5 Usage Testing*

*Robert M. Patton et al.*
Robert M. Patton [23] has used the usage models that depict the usage scenarios of the system. They are used in test planning, to generate a sample of test cases that represent usage scenarios, and to test results. In system testing, determining the nature and the location of the errors can be difficult which later on can be the problem for the developers to fix the errors. The system testing contains only small information from the usage scenarios of the system. Due to the limited information, generalizing testing results could be difficult.

To solve these problems, GA accepts the domain data generated by the usage model and the results of system test as two inputs. A set of test cases are the initial population generated from a usage model. Each individual in the population represents a single test case. The two objectives were taken to determine the fitness of individuals. The first objective is 'Likelihood of occurrence' which represents the possible usage scenarios of what the user will do with the system. The second objective is 'failure intensity' which is the capability of an individual to exhibit failures or problems in the system. The individuals maximizing these two objectives are selected for mating.

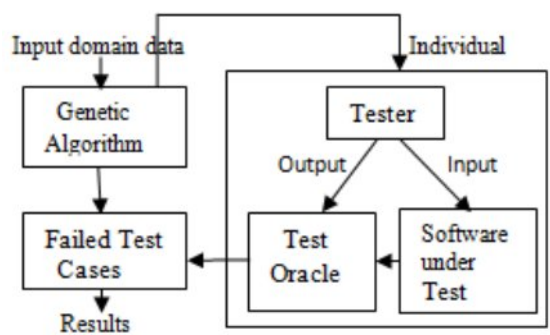

Fig. 5 R. Patton's GA approach to Focused software usage testing [23]

As shown in Fig 5, each individual represents a single test case and is sent to the tester and is then applied to SUT. The SUT processes this input and generates the output that is later analysed by the Test Oracle. The Test Oracle will then determine if the output is correct or incorrect or if the SUT failed or crashed. The GA uses this result along with the likelihood that it would occur as defined by the usage model for determining the overall fitness of the individual.

*Discussion*

The R. Patton's strategy shows that GA helps to identify failures that are more severe and are likely to cause faults in the software.

3.2.6 GUI testing

*Abdul Rauf et. al.*
In this work [1], GA is used to apply coverage criteria on GUI (Graphical User Interface). GUI is event based testing where the test cases consist of GUI events. GA is used to find optimized test suite for GUI testing. GUI testing is divided into three phases:-

1. Test Data Generation

2. Path Coverage Analysis

3. Optimization of Test Path

The Notepad, MS Word and Word Pad are used for the event based test data generation.

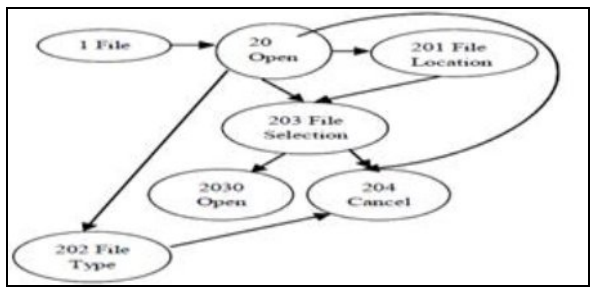

Fig. 6 Path generation for Open in Notepad [1]

In Fig 6, events generation while opening a file is shown. The nodes represent the objects in a notepad like File, Print, Edit and the sequence of operations between the nodes is shown as path between them. The test paths are optimized using GA. The size of chromosome in GA is measured as maximum length of test path in a Notepad. The fitness function is defined as number of paths covered by chromosome shown in equation 19.

$$Accuracy \quad \frac{TestPathsCoveredbyChromosome}{TotalNumberofChromosome} \quad (19)$$

By applying crossover and mutation operator, new off springs are obtained having higher fitness value.

*Discussion*
The results show 85% coverage after 500 generation. Since the probability of crossover and mutation is unknown, the results might be superior if the crossover and mutation rate is changed.

### 3.3 Applications of GA in Gray box testing

In gray box testing, test cases are designed using both black box testing and white box testing. The software is tested against its specification but using some knowledge of internal working [8] [30].

### 3.3.1 User Session Testing

*Xuang Peng et. al.*
In [30], Xuan Peng et. al. used the user session data in their request dependence graph (RDG) to generate test cases by applying GA. The structural analysis of web application is done using RDG construction. The request dependence is shown as the relationship between the components or nodes of the web application. The edge represents the request dependence between two pages. The request labelled in the graph is formatted as: - "GET/POST PAGE < P1, P2,......,Pn > where P1, P2,....., Pn are the possible parameters in the request.

All the parameter- value pair of requests is enumerated without values. The RDG and their corresponding labelling are shown in Fig 7.

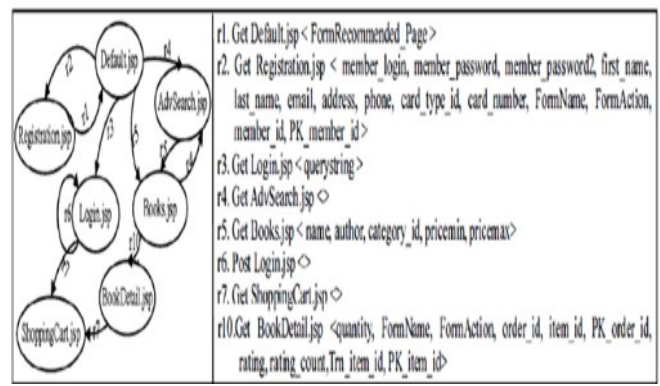

Fig. 7 Partial Request Dependence Graph and the Labelled Requests [30]

The test cases should cover as many relationships as possible in RDG. The user session data is taken as the initial population in GA. A gene is encoded as combination of requests and pages. User session is identified as a sequence of requests and parameter values to describe the user's requests for web services. Each session is represented as transition relation. e.g. "Request -> Page -> Request -> ..............->Page -> Request". A chromosome is encoded as a transition relationship between page and request. The fitness value is computed as: -

$$Fitness \quad (\infty *|CDTR| + |CLTR|)/(\infty *|DTR| + |LTR|) \quad (20)$$

Where, CDTR is number of data dependence transitions covered in the chromosome, CLTR is number of link dependence transitions covered in the chromosome. DTR is number of data dependence and LTR is number of link dependence transition relations in the web application.

Discussion

Results show that user session (US) – RDG performed well for web application testing. Test suite reduction and fault detection results are quite satisfactory.

A comparative study of all existing techniques discussed in our work is shown in Table 3 which shows the values of GA parameter used in different types of software testing.

Table 3: GA Parameters used in different types of software testing (C.R = Crossover Rate, M. R = Mutation Rate, N.O.G = Number of Generation)

| Author | Testing Technique | C. R | M. R | Crossover Method | Mutation Method | Selection | Initial Pop. size | N.O.G | Encoding | Result |
|---|---|---|---|---|---|---|---|---|---|---|
| D. J Berndt, 2005 | High Volume Testing | -- | -- | -- | -- | -- | -- | 500 | Real numbers | GA is being mined with decision tree to generate test cases. To explore the GA, an autonomic vehicle simulation is being developed. Robots are built with limited resources and seeded with errors. On long term execution, robot simulation shows the promising results in the high volume testing. |
| Doungsa | Gray-box testing | 0.4 | 0.3 | Single one point | Random | Random | 6 | 2 | Value encoding | GA succesfully generates test data from the UML state diagram. |
| Dr. Velur Rajappa, 2008 | System testing or network testing | 50% | 50% | Single one point | -- | Tournament selection | 250 | -- | Tree encoding | The solution is proposed using graph theory and genetic algorithm. |
| Francisca Emanuelle [35], 2006 | Functional Testing | 80% | 1% | Point Crossover | Flip | Random | 150 | 50 | -- | Test plans showing highest inconsistency of application. |
| Jose Carlos, 2008 | Unit testing | 0.1, 0.8, 0.33 | 0.1, 0.8, 0.34 | Random point | > 7.5 average generations (Combination of C.R and M. R and r =[0.1, 0.8, 0.33] | Tournament selection | -- | -- | Tree encoding | Weighted CFG is used where the weights of CFG are dynamically revaluated to determine the qulaity of test cases. |
| Maha Alzabidi, 2009 | Path testing | 1.0 (Tri class program) and Max-Min Program) | 0.005 ( Tri class) and 0.05 for Max- Min program | Double point for Tri class program and Single -point for Max- Min | Flip | Random and roulette wheel | 500 for Tri - class and 500 for Max- Min | 50 for Tri - class and 10 for Max- Min | Binary | 1. In double point there are good chances to double exchange the generation than single point crossover. 2. Mutation o.005 gave better results 3. Generation of next population according to their fitness value generates better offspring than random selection. |
| Mark Last, 2006 | Black Box | Adaptive | 0.01 | One point | Flip | Random | 100 | 200 | Binary | FAexGA is efficient than simple GA and GA with varying population size in terms of probability of finding an error in tested software, faster rate of finding error and number of distinct solution. |
| Moheb R. Girgis, 2005 | Data flow testing | 0.8 | 0.15 | -- | Flip | Random selection and Roulette wheel selection | 4 | -- | Binary | Propsed Random selection technique requies less number of generation to cover data flow dependencies than roulette wheel selection. |
| Nirmal Kumar Gupta, 2008 | Unit testing | -- | -- | -- | Random | Random or execution traces | -- | -- | Tree encoding | Test case generation for java classes and encoding and decoding of test program into changeable data structure. |
| P. R Srivastava, 2009 | Path Testing | For r = [0,1], r < 0.8 | For r = [0, 1], r < 0.3 | Pairwise | Flip | Random | 4 | 3 | Binary | Increase in Testing efficiency by testing critical path in CFG |
| R. Krishnamoorthy, 2009 | Regression testing | for r = [0,1], r < user value | 1% | Random point | Random | Roulette Wheel | 60 | 25 | Real numbers | Using GA, the proposed time aware coverage based prioritization technique shows 120% improvement in APFD over other prioritization. |
| Robert M. Patton, 2003 | Usage testing | -- | -- | One point | Random One - point | Fitness proportionate | 100 | 30 | Real numbers | GA is used in focused sofware usage testing by identifying the nature and locations of errors thereby improving the quality of software and efficiency of debugging activity. |
| Ruilian Zhao , 2008 | Black Box | 0.8 | 0.15 | -- | -- | Roulette Wheel | -- | > 500 | Real | 1. Improved GA is better than faster evolutionary speed 2. Approach can generate test cases with high efficiency. |
| Stefan Wappler, 2006 | Unit Testing | -- | -- | Point crossover | Point mutation & Real Mutation | Stochastic universal Sampling & Tournament Selection | 10 | <10 | -- | Full branch coverage achieved for all test objects |
| Vahid Garousi, 2008 | Stress Testing | 50% & 70% | -- | -- | -- | -- | -- | < 100 , for C.R = 50% and > 100 for C. R = 70 % | -- | 1. For C. R = 70%, fitness value increased by 80 % 2. GA can reach max. plateau even size of a component in SUT is large  3. Maximum search time delays the convergence across GA to find the best chromosome. |

## 4. Conclusion and Future Work

In this paper, applications of GA in different types of software testing are discussed. The GA is also used with fuzzy as well as in the neural networks in different types of testing. It is found that by using GA, the results and the performance of testing can be improved. Our future work will involve applying GA for regression testing in web based applications. In future, we plan to use GA along with other soft computing techniques like fuzzy logic or neural networks for test case generation from UML diagrams. We also plan to use GA in integration testing for finding optimal test order.